\documentclass[twocolumn,prl,showpacs]{revtex4}
\usepackage{amsmath}
\usepackage{amssymb}
\usepackage{graphicx}
\usepackage{esint}

\begin{document}

\title{Optimal Implementations for Reliable Circadian Clocks}

\author{Yoshihiko Hasegawa$^{1}$}

\email[Corresponding author: ]{yoshihiko.hasegawa@gmail.com}
\email[Present address: Department of Information and Communication Engineering, Graduate School of Information Science and Technology, The University of Tokyo, Tokyo 113-8656, Japan]{}

\author{Masanori Arita$^{2,3}$}

\affiliation{$^{1}$Department of Biophysics and Biochemistry, Graduate School
of Science, The University of Tokyo, Tokyo 113-0033, Japan;}

\affiliation{$^{2}$Center for Information Biology, National Institute of Genetics,
Shizuoka 411-8540, Japan;}

\affiliation{$^{3}$RIKEN Center for Sustainable Resource Science, Kanagawa 230-0045,
Japan.}
\begin{abstract}
Circadian rhythms are acquired through evolution to increase the chances
for survival through synchronizing with the daylight cycle. Reliable
synchronization is realized through two trade-off properties: regularity
to keep time precisely, and entrainability to synchronize the internal
time with daylight. We found by using a phase model with multiple
inputs that achieving the maximal limit of regularity and entrainability
entails many inherent features of the circadian mechanism. At the
molecular level, we demonstrate the role sharing of two light inputs,
phase advance and delay, as is well observed in mammals. At the behavioral
level, the optimal phase-response curve inevitably contains a dead
zone, a time during which light pulses neither advance nor delay the
clock. We reproduce the results of phase-controlling experiments entrained
by two types of periodic light pulses. Our results indicate that circadian
clocks are designed optimally for reliable clockwork through evolution. 
\end{abstract}


\maketitle
Many terrestrial species, from cyanobacteria through to humans, adapt
to sunlight and have acquired circadian oscillatory systems. Although
their molecular implementation is species dependent \cite{Young:2001:GeneticsCirc},
all exhibit a regular rhythm of 24-h that can be entrained by light
input. Two fundamental properties are necessary for circadian clocks
\cite{Johnson:2011:CyanoCircadian}, i.e. \emph{regularity} to keep
time precisely \cite{Vilar:2002:NoiseResistGeneOsc,Gonze:2002:RobustCircadian,Herzon:2004:CircadianPrecision}
and \emph{entrainability} to adjust the internal time through light
stimuli \cite{Gonze:2000:Entrainment,Roenneberg:2003:ArtEntrain,Golombek:2010:CircadianEntrainment}.
It is not easy to maximize entrainability and regularity simultaneously;
higher entrainability implies more vulnerability to noise, whereas
higher regularity signifies less flexibility to entrainment. We studied
an optimal phase-response curve (PRC) problem for a one-input pathway
in \cite{Hasegawa:2013:OptimalPRC}. In the present Letter, we generalize
the calculations of \cite{Hasegawa:2013:OptimalPRC} and formalize
the optimal implementation problem for multiple input pathways without
relying on specific oscillator models \cite{Roenneberg:2003:ArtEntrain,Troein:2009:ComplexClocks,Troein:2011:TauriClock}.
We show that the simultaneous maximization of entrainability and
regularity entails several inherent properties of circadian clocks.
At the molecular level, we rationalize the role sharing of multiple
input pathways which was reported in murine circadian clocks \cite{Albrecht:2001:mPer1_mPer2,Steinlechner:2002:MutantLL}.
At the behavioral level, we rationalize a time period during which
the time is neither advanced nor delayed by light in an optimal clock
\cite{Refinetti:2005:CircBook}. Our theory can also explain different
gene expression patterns when entrained by two different light pulses
\cite{Schwartz:2011:PerRoles}. In this study we investigate the
optimal implementations for multiple input pathways to achieve maximal
limit of entrainability and regularity. Although all circadian clocks
transmit light signals through multiple pathways \cite{Roenneberg:2003:ArtEntrain,Troein:2009:ComplexClocks,Troein:2011:TauriClock},
entrainment problems in multiple pathways have been little studied. 

Circadian clocks keep time regularly under molecular noise \cite{Vilar:2002:NoiseResistGeneOsc,Gonze:2002:RobustCircadian,Herzon:2004:CircadianPrecision}.
It is also entrainable by periodic light stimuli. To consider these
effects, we use an $N$-dimensional Langevin equation with respect
to $x_{i}$, which is the concentration of $i$th molecular species:
$\dot{x}_{i}=F_{i}(\boldsymbol{x};\rho)+Q_{i}(\boldsymbol{x})\xi_{i}(t)$
where $F_{i}(\boldsymbol{x};\rho)$ is the $i$th reaction rate ($i=1,2,..,N$),
$Q_{i}(\boldsymbol{x})$ is a multiplicative noise term, and $\xi_{i}(t)$
is white Gaussian noise with $\left\langle \xi_{i}(t)\xi_{j}(t^{\prime})\right\rangle =2\delta_{ij}\delta(t-t^{\prime})$.
To incorporate the effect of light, we introduce a light-sensitive
parameter $\rho$ which is perturbed as $\rho\rightarrow\rho+d\rho$
when stimulated by light. We quantify regularity by the temporal variance
in the oscillation. From \cite{Hasegawa:2013:OptimalPRC}, the period
variance is known to be $\mathcal{V}_{T}\simeq T^{3}(4\pi^{3})^{-1}\int_{0}^{2\pi}\sum_{i=1}^{N}U_{i}(\theta)^{2}Q_{i}(\theta)^{2}d\theta$,
where $T$ is the period of the unperturbed oscillator, $\boldsymbol{U}(\phi)=(U_{1}(\phi),\cdots,U_{N}(\phi))$
is the infinitesimal PRC (iPRC) defined by $\boldsymbol{U}(\phi)=\left.\nabla_{\boldsymbol{x}}\phi\right|_{\boldsymbol{x}=\boldsymbol{x}_{\mathrm{LC}}(\phi)}$
and $\boldsymbol{x}_{\mathrm{LC}}(\phi)$ is a point on the limit-cycle
trajectory at phase $\phi\in[0,2\pi)$. Entrainability is quantified
as the width of the Arnold tongue. The phase dynamics with periodic
input signals are described by a tilted periodic potential. If stable
points exist in the potential, the oscillator can be entrained by
input signals \cite{Kuramoto:2003:OscBook}. Thus we can discuss entrainability
without considering noise because the existence of stable points does
not depend on noise. Therefore we set $Q_{i}(\boldsymbol{x})=0$.
Let $p(\omega t)$ be an input signal with angular frequency $\omega$.
For a weak input signal $d\rho=\chi p(\omega t)$ ($\chi\ge0$ is
the signal strength), we obtain $\dot{\phi}=\Omega+\chi Z(\phi)p(\omega t)$
($\phi\in[0,2\pi)$ is the phase and $\Omega=2\pi/T$) where 
\begin{equation}
Z(\phi)=\sum_{i=1}^{N}Z_{i}(\phi),\hspace{1em}Z_{i}(\phi)=\frac{\partial F_{i}(\phi;\rho)}{\partial\rho}U_{i}(\phi),\label{eq:PRC_Z}
\end{equation}
with $F_{i}(\phi;\rho)=F_{i}(\boldsymbol{x}_{\mathrm{LC}}(\phi);\rho)$.
$Z_{i}(\phi)$ quantifies phase shift due to the perturbation of $\rho$
in the $i$th coordinate. We hereafter refer to $Z(\phi)$ as the
\emph{parametric PRC} (pPRC) \cite{Taylor:2008:Sensitivity,Hasegawa:2013:OptimalPRC}.
The dynamics of a slow variable defined by $\psi=\phi-\omega t$ obeys
\begin{equation}
\dot{\psi}=\Omega-\omega+\chi\Theta(\psi),\label{eq:psi_dynamics}
\end{equation}
where $\Theta(\psi)=(2\pi)^{-1}\int_{0}^{2\pi}Z(\psi+\theta)p(\theta)d\theta$.
The width of the Arnold tongue is given by $\chi\mathcal{E}$, where
$\mathcal{E}=(2\pi)^{-1}\int_{0}^{2\pi}Z(\theta)\left\{ p(\theta-\psi_{M})-p(\theta-\psi_{m})\right\} d\theta$
with $\psi_{M}=\mathrm{argmax}_{\psi}\Theta(\psi)$ and $\psi_{m}=\mathrm{argmin}_{\psi}\Theta(\psi)$
\cite{Hasegawa:2012:GeneOsc,Hasegawa:2013:OptimalPRC}. Optimal circadian
clocks are derived as optimal PRCs $U_{i}(\phi)$ which were obtained
with the variational method by maximizing the entrainability $\mathcal{E}$
subject to constant variance $\mathcal{V}_{T}=\sigma_{T}^{2}$: $U_{i}(\phi)=\pi^{2}T^{-3}\lambda^{-1}Q_{i}(\phi)^{-2}\left\{ p(\phi-\psi_{M})-p(\phi-\psi_{m})\right\} \partial_{\rho}F_{i}(\phi;\rho)$
($\lambda$ is a Lagrange multiplier \cite{SI})
\cite{Hasegawa:2013:OptimalPRC}.

We generalize the above scheme to multiple input signals. Suppose
there are $M$ input pathways and parameters $\rho_{1},\rho_{2},\cdots,\rho_{M}$
are affected as $\rho_{i}\rightarrow\rho_{i}+d\rho_{i}$ under light
stimuli. By introducing an auxiliary scaling parameter $s_{i}$ defined
by $d\rho_{i}=s_{i}d\rho$, each parameter perturbation is reparameterized
as 
\begin{equation}
\rho_{i}\rightarrow\rho_{i}+d\rho_{i}\Rightarrow\tilde{\rho}_{i}+s_{i}\rho\rightarrow\tilde{\rho}_{i}+s_{i}(\rho+d\rho),\label{eq:reparam}
\end{equation}
where $\tilde{\rho}_{i}=\rho_{i}-s_{i}\rho$. Through this reparameterization,
the multiple perturbations can be reduced to a single parameter case
with respect to $\rho$. In this instance, $\rho$ in Eq.~\eqref{eq:reparam}
is a hypothetical parameter and has no correspondence to actual reaction
rates.

\begin{figure}
\begin{centering}
\includegraphics[width=8cm]{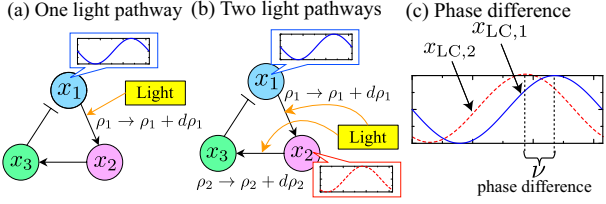} 
\par\end{centering}

\caption{Examples of (a) one-light input pathway and (b) two-light input pathway
cases. The insets describe the time course of molecular species that
are affected by light stimuli. In (b) the light-sensitive molecular
species correspond to $x_{1}$ and $x_{2}$ that exhibit peaks at
different phase (this difference is given by $\nu$). (c) The phase
difference $\nu$ is defined by the difference between $k_{1}$th
and $k_{2}$th molecular species. \label{fig:illustrations}}
\end{figure}

Light stimuli generally affect the rate constants, but the detailed
mechanisms vary between organisms \cite{Johnsson:2007:CircLightReview}.
Our model can cover these different entrainment mechanisms. We first
explain a simple case (Fig.~\ref{fig:illustrations}(a)), and then
generalize it. In Fig.~\ref{fig:illustrations}(a), we assume that
light stimuli enhance the synthesis rate of $x_{2}$, where the dynamics
of $x_{2}$ can be described by $\dot{x}_{2}=\rho_{\mathrm{syn}}x_{1}+\rho_{\mathrm{deg}}x_{2}$
where $\rho_{\mathrm{syn}}>0$ and $\rho_{\mathrm{deg}}<0$ are the
synthesis and degradation rates, respectively. Because the synthesis
rate increases when stimulated by light, by taking $\rho=\rho_{\mathrm{syn}}$
($\rho$ is the light-sensitive parameter), the rate equation can
be divided into $\rho$-dependent and $\rho$-independent parts $\dot{x}_{2}=F_{2}(\boldsymbol{x};\rho)=\rho_{\mathrm{deg}}x_{2}+\rho x_{1}=\tilde{F}_{2}(\boldsymbol{x})+\rho x_{1}$,
where $\tilde{F}_{2}(\boldsymbol{x})$ denotes terms not including
$\rho$ in $F_{2}(\boldsymbol{x};\rho)$. We next consider a generic
case with two-input pathways ($M=2$). Let us assume that light stimuli
affect parameters $\rho_{1}$ and $\rho_{2}$ (e.g. translation, transcription
or degradation rate) which depend on $k_{1}$th and $k_{2}$th molecular
species and affect $j_{1}$th and $j_{2}$th species, respectively
(e.g. $k_{1}=1$, $k_{2}=2$, $j_{1}=2$ and $j_{2}=3$ in Fig.~\ref{fig:illustrations}(b)).
The rate equations $F_{j_{i}}(\boldsymbol{x};\rho)$ of $j_{1}$th
and $j_{2}$th are described as 
\begin{align}
\dot{x}_{j_{i}} & =\tilde{F}_{j_{i}}(\boldsymbol{x})+\rho_{i}x_{k_{i}}=\underbrace{\tilde{F}_{j_{i}}(\boldsymbol{x})+\tilde{\rho}_{i}x_{k_{i}}}+\underbrace{s_{i}\rho x_{k_{i}}},\label{eq:j_rate_equation}\\
 & \qquad\qquad\qquad\qquad\quad\rho\mbox{-}\mathrm{independent}\quad\rho\mbox{-}\mathrm{dependent}\nonumber 
\end{align}
where $i=1,2$ and $\tilde{F}_{j_{i}}(\boldsymbol{x})$ denotes terms
not including $\rho_{i}$ in $F_{j_{i}}(\boldsymbol{x};\rho)$. From
Eq.~\eqref{eq:PRC_Z}, we see that PRCs are only concerned with the
derivative of $F_{j_{i}}(\boldsymbol{x};\rho)$ with respect to $\rho$,
which is specifically given by $\partial_{\rho}F_{j_{i}}(\boldsymbol{x};\rho)=s_{i}x_{k_{i}}$.
Consequently, the $\rho$-independent part in Eq.~\eqref{eq:j_rate_equation}
plays no role in the formation of optimal PRCs. We approximate $x_{k}$
by the $k$th coordinate of $\boldsymbol{x}_{\mathrm{LC}}$, denoted
as $x_{\mathrm{LC},k}(\phi)$ in Eq.~\eqref{eq:j_rate_equation}.
Because $x_{\mathrm{LC},k}(\phi)$ oscillates, we model it with a
sine curve. $k_{1}$th and $k_{2}$th molecular species may peak at
different times as shown in Fig.~\ref{fig:illustrations}(b), where
the time courses of the $k_{1}$th and $k_{2}$th molecular species
is described in the insets by solid and dashed lines, respectively.
Therefore, the time course is approximated by $x_{\mathrm{LC},k_{i}}(\phi)=1-\alpha_{i}\sin\left(\phi+u_{k_{i}}\right)$
for $i=1,2$, where $u_{k_{1}}$ and $u_{k_{2}}$ are initial phases
of $k_{1}$th and $k_{2}$th molecular species, the amplitude $\alpha_{i}$
being assumed as identical $\alpha_{1}=\alpha_{2}=\alpha$ ($0\le\alpha\le1$).
Let $\nu$ be the phase difference between $k_{1}$th and $k_{2}$th
molecular species, i.e. $\nu=u_{k_{2}}-u_{k_{1}}$ (Fig.~\ref{fig:illustrations}(c)).
This parameter plays an important role in optimization.

Circadian clocks are entrained by sunlight whose intensity is determined
by 24-h periodic solar irradiance. The solar irradiance $I$ with
respect to the zenith angle $\vartheta$ is given by $I=I_{0}\cos\vartheta$,
where $I_{0}$ is the irradiance at $\vartheta=0$, and $I$ vanishes
when the sun is below the horizon \cite{Hartmann:1994:PhysClimate}.
Thus we approximate its time course by $p(\omega t)=\sin(\omega t)$
for $0\le\mod(\omega t,2\pi)<\pi$ (day) and $p(\omega t)=0$ for
$\pi\le\mod(\omega t,2\pi)<2\pi$ (night) with $\mod(x,y)=x\,\mod\, y$,
which we call \emph{solar radiation input}~\cite{Hasegawa:2013:OptimalPRC}.

\begin{figure}
\begin{centering}
\includegraphics[width=8.5cm]{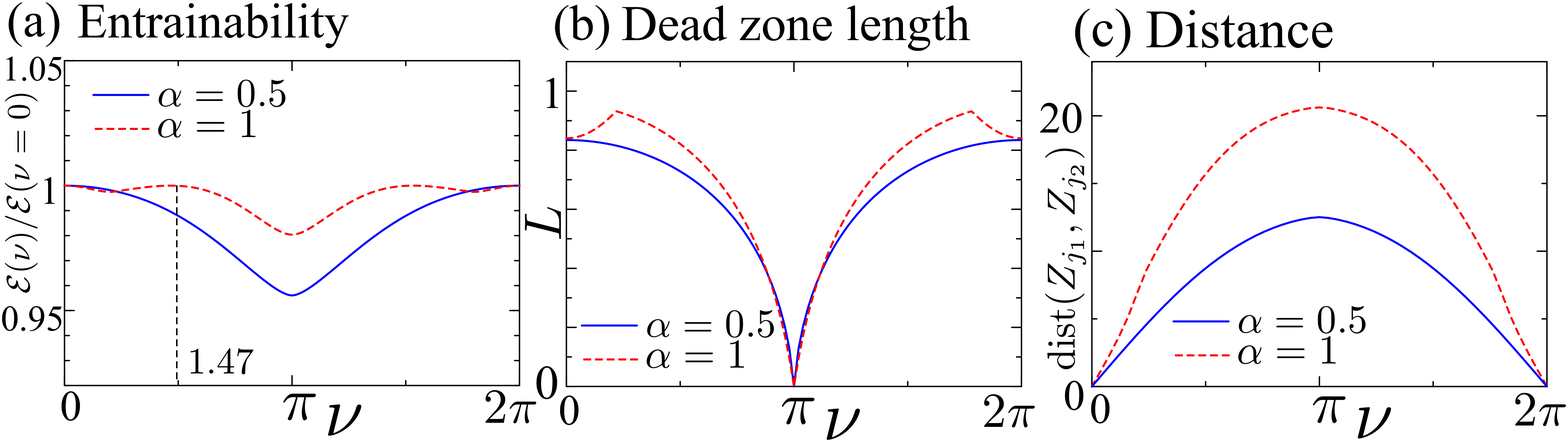} 
\par\end{centering}

\caption{(a) Normalized entrainability as a function of $\nu$ for $\alpha=0.5$
(solid line) and $\alpha=1$ (dashed line). The value is normalized
by the entrainability at $\nu=0$ (i.e. $\mathcal{E}(\nu)/\mathcal{E}(\nu=0)$).
For $\alpha=1$, maximal is achieved at $\nu=0$ and $1.47$. (b)
Dead zone length $L$ as a function of the phase difference $\nu$
for $\alpha=0.5$ (solid line) and $\alpha=1$ (dashed line). (c)
Distance between $Z_{j_{1}}(\phi)$ and $Z_{j_{2}}(\phi)$ as a function
of $\nu$ for $\alpha=0.5$ (solid line) and $\alpha=1$ (dashed line).
\label{fig:diff_dependence}}
\end{figure}

\begin{figure}
\begin{centering}
\includegraphics[width=7cm]{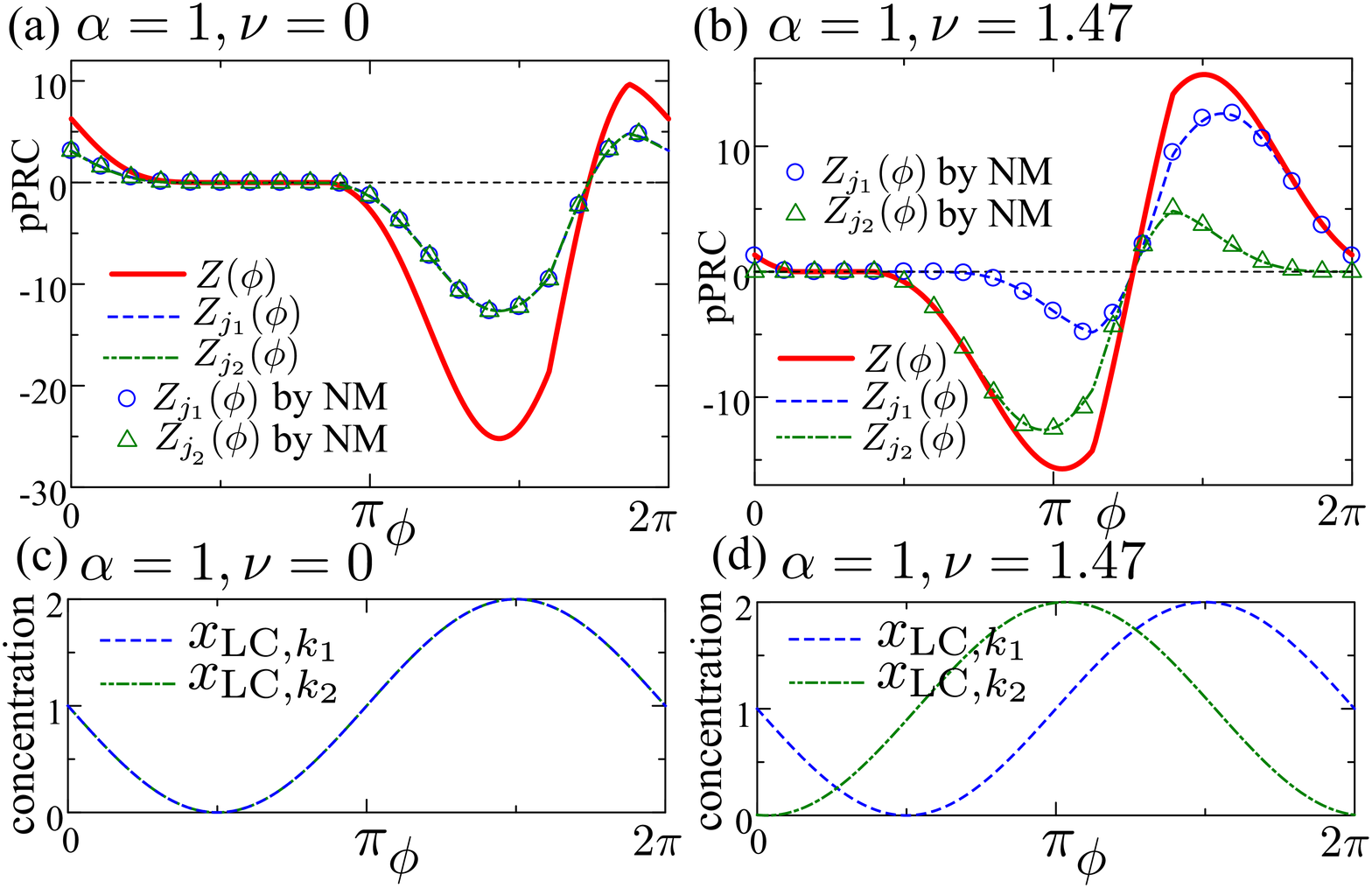} 
\par\end{centering}

\caption{(a)--(b) Optimal PRCs ($Z(\phi)$, $Z_{j_{1}}(\phi)$, and $Z_{j_{2}}(\phi)$)
for (a) $\alpha=1$ and $\nu=0$ and (b) $\alpha=1$ and $\nu=1.47$.
In (a) and (b), $Z(\phi)$, $Z_{j_{1}}(\phi)$ and $Z_{j_{2}}(\phi)$
obtained with the variational method, are plotted with thick solid
lines, dashed lines and dot-dash lines, respectively. $Z_{j_{1}}(\phi)$
and $Z_{j_{2}}(\phi)$ obtained with a numerical method (NM) are plotted
with circles and triangles, respectively \cite{SI}.
In (a), PRCs that are symmetric with respect to the horizontal axis
or $\phi=3\pi/2$ are also optimal. In (b), PRCs that are symmetric
with respect to the horizontal axis are also optimal. (c)--(d) Time
course of the molecular species concentration for (c) $\alpha=1$
and $\nu=0$ and (d) $\alpha=1$ and $\nu=1.47$. Two species $x_{\mathrm{LC},k_{1}}$
and $x_{\mathrm{LC},k_{2}}$ are plotted with dashed and dot-dash
lines. In (a) and (c), dashed and dot-dashed lines are indistinguishable.
\label{fig:optimal_PRC_two}}
\end{figure}

Our model starts from the two-input case ($M=2$). The noise term
was introduced only for input molecules ($Q_{j_{1}}(\phi)=\sqrt{q}$,
$Q_{j_{2}}(\phi)=\sqrt{q}$ and $Q_{i}(\phi)=0$ otherwise, where
$q$ is the noise intensity), and we set $u_{k_{1}}=0$ (i.e. $\nu=u_{k_{2}}$),
$T=1$, $\sigma_{T}^{2}=1$, and $q=1$ without loss of generality.
We then need to specify additional parameters $s_{1}$ and $s_{2}$
(Eq.~\eqref{eq:reparam}) that determine the strength of the input
signal relative to $\rho$. If we are concerned with pPRCs (experimentally
observed PRCs) only, the sign of $s_{i}$ (positive or negative) does
not play any role, because $s_{i}$ is squared in the optimal pPRCs.
We set $s_{1}=s_{2}=1$ for simplicity; this corresponds to assuming
equal weight for each light input pathway.

Figure ~\ref{fig:diff_dependence}(a) shows the $\nu$ dependence
of maximal entrainability for $\alpha=0.5$ (solid line) and $\alpha=1$
(dashed line). For $\alpha=0.5$, the maximum was achieved at $\nu=0$
where no phase difference existed. However, upon increasing $\alpha$
to $1$, maximal entrainability was achieved at two points, $\nu=0$
and $\nu=1.47$ (Figs.~\ref{fig:optimal_PRC_two}(a) and (b)). Interestingly,
optimality can be attained in the presence of phase difference. From
Eq.~\eqref{eq:PRC_Z}, we divide the pPRC $Z(\phi)$ into contributions
from two input pathways $Z(\phi)=Z_{j_{1}}(\phi)+Z_{j_{2}}(\phi)$,
where $Z_{j_{1}}(\phi)$ and $Z_{j_{2}}(\phi)$ quantify the phase
shift produced by the 1st and 2nd input pathways, respectively. Optimal
$Z(\phi)$, $Z_{j_{1}}(\phi)$, and $Z_{j_{2}}(\phi)$ for the two-input
case are plotted in Figs.~\ref{fig:optimal_PRC_two}(a) and (b).
Figures~\ref{fig:optimal_PRC_two}(c) and (d) are the corresponding
time course of the $k_{1}$th and $k_{2}$th molecular species concentration.
We see that optimal PRCs $Z(\phi)$ in Figs.~\ref{fig:optimal_PRC_two}(a)
and (b) are very similar to experimentally observed PRCs in which
there is a dead zone, a time during which light neither advanced nor
delayed the clock ($1\lesssim\phi\lesssim2$ in Figs.~\ref{fig:optimal_PRC_two}(a)
and (b)). Intriguingly, experimental studies in different species
reported the existence of the dead zone \cite{Refinetti:2005:CircBook}.
Although cases with $\nu=0$ and $\nu=1.47$ achieved the same entrainability,
$Z(\phi)$ for $\nu=0$ is asymmetric with respect to horizontal axis,
which entails an asymmetric Arnold tongue. Thus for a symmetric Arnold
tongue, only $\nu=1.47$ can achieve maximal entrainability. We calculated
the $\nu$ dependence of the dead zone length $L$ (length of null
parts in PRCs \cite{SI}) in Fig.~\ref{fig:diff_dependence}(b)
for $\alpha=0.5$ (solid line) and $\alpha=1.0$ (dashed line). In
Fig.~\ref{fig:diff_dependence}(b), $L$ quickly diminishes around
$\nu=\pi$, showing that a dead zone always appears in optimal PRCs
except for a singular point $\nu=\pi$. 

The fundamental difference of $\nu=1.47$ from $\nu=0$ is the role
sharing of two PRCs $Z_{j_{1}}$ and $Z_{j_{2}}$ (Fig.~\ref{fig:optimal_PRC_two}(b)).
$Z_{j_{1}}$ is responsible for the phase advance and $Z_{j_{2}}$
for the delay (the positive part of $Z_{j_{1}}$ is larger than the
negative part and vice versa). This effect was observed for all $\nu$
values except $\nu=0$, as shown below. We quantify the distance between
$Z_{j_{1}}$ and $Z_{j_{2}}$ by $\mathrm{dist}(Z_{j_{1}},Z_{j_{2}})=\sqrt{\int_{0}^{2\pi}\left\{ Z_{j_{1}}(\theta)-Z_{j_{2}}(\theta)\right\} ^{2}d\theta}$,
which becomes larger when the two PRCs play more compensatory roles.
The distance calculated as a function of $\nu$ for $\alpha=0.5$
(solid line) and $\alpha=1$ (dashed line) is shown in Fig.~\ref{fig:diff_dependence}(c)
where the distance is maximal exactly at $\nu=\pi$. When a phase
difference ($\nu>0$) exists, this role-sharing between two-input
pathways always yields a synchronization advantage. There is experimental
evidence for advance and delay roles of \textit{Per}1 and \textit{Per}2,
respectively, in mice \cite{Albrecht:2001:mPer1_mPer2}. In this regard,
Ref.~\cite{Steinlechner:2002:MutantLL} observed a period dependence
of \emph{Per}1 and \emph{Per}2 knockout mutants on the intensity of
constant light. We can reproduce this result with optimal PRCs as
follows. Note that entrainability is maximal at $\nu=1.47$ for $\alpha=1$
(Fig.~\ref{fig:optimal_PRC_two}(b)). Consequently, we set the clock
parameters to $\alpha=1$ and $\nu=1.47$. Under a constant light
condition, the input signal is modeled by $d\rho=\chi p(t)$, where
$p(t)=1$. By integrating the phase equation $\dot{\phi}=\Omega+\chi Z(\phi)p(t)$
from $t=0$ to $t=T$ where $\phi(t=0)=0$, the phase at time $T$
with input strength $\chi$ is given by $\phi(T;\chi)=2\pi+(2\pi)^{-1}T\chi\int_{0}^{2\pi}Z(\theta)d\theta$.
For weak $\chi$, the period $T_{\chi}$, which is the period under
a constant light condition, is approximated by $T_{\chi}/T\simeq\phi(T;0)/\phi(T;\chi)\simeq1-T\chi(4\pi^{2})^{-1}\int_{0}^{2\pi}Z(\theta)d\theta$ \cite{SI,Taylor:2008:Sensitivity}. 
Assuming $x_{j_{1}}=[Per1]$ and $x_{j_{2}}=[Per2]$, we simulated
\emph{Per}1 and \emph{Per}2 mutants by setting $Z(\phi)=Z_{j_{2}}(\phi)$
and $Z(\phi)=Z_{j_{1}}(\phi)$, respectively. When increasing the
intensity $\chi$ of constant light, the period ratio $T_{\chi}/T$
increases for \emph{Per}1 mutant and decreases for \emph{Per}2 mutant.
This result agrees with the experimental evidence (Fig.~2 in \cite{Steinlechner:2002:MutantLL}).

\begin{figure}
\begin{centering}
\includegraphics[width=8.5cm]{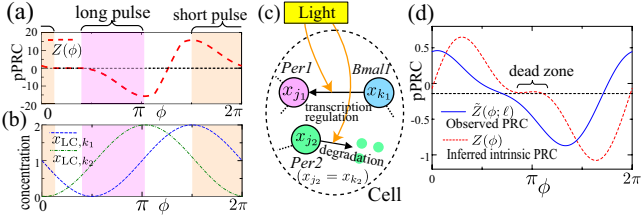} 
\par\end{centering}

\caption{(a)--(c) Theoretical reproduction of a light entrainment experiment
involving hamsters \cite{Schwartz:2011:PerRoles}. (a) Optimal pPRC
$Z(\phi)$ of the two-input case $M=2$ with $\alpha=1$ and $\nu=1.47$
(Fig.~\ref{fig:optimal_PRC_two}(b)). (b) The time course of the
molecular species concentration of $x_{\mathrm{LC},k_{1}}$ (dashed
lines) and $x_{\mathrm{LC},k_{2}}$ (dot-dash lines) for $\alpha=1$
and $\nu=1.47$. (c) Molecular implementation of the murine light
entrainment mechanism. The model can be described by Eq.~\eqref{eq:j_rate_equation}
with $x_{j_{1}}=[Per1]$, $x_{k_{1}}=[Bmal1]$, and $x_{j_{2}}=x_{k_{2}}=[Per2]$.
(d) Observed (solid line) and inferred-intrinsic ($N_{\mu}=2$, $\chi=1$;
dashed line) pPRCs of humans as a function of the onset of pulses.
Observed pPRC is brought from Ref.~\cite{Khalsa:2003:HumanPRC},
which measured pPRC with $\ell=6.7$-h light pulses. The horizontal
dashed line indicates anticipated phase delay \cite{Khalsa:2003:HumanPRC}.
\label{fig:optimal_pathway}}
\end{figure}

Our model can further suggest insights into the molecular mechanism
of the clock. In hamsters, Schwartz \textit{et al.} \cite{Schwartz:2011:PerRoles}
reported different gene expressions of \textit{Per}1 and \textit{Per}2
when entrained by two types of periodic light pulses that have short
(23.33-h) and long (24.67-h) periods. Let us reproduce Schwartz's
experiment in our optimization framework with two inputs. We again
set $\alpha=1$ and $\nu=1.47$ and assume $x_{j_{1}}=[Per1]$ and
$x_{j_{2}}=[Per2]$. Ref.~\cite{Schwartz:2011:PerRoles} applied
a periodic light pulse of $1$-h duration, which we modeled with a
periodic $\delta$-function 
\begin{equation}
p(\omega t)=2\pi\delta(\mathrm{mod}(\omega t,2\pi)),\label{eq:periodic_pulse}
\end{equation}
where a factor $2\pi$ ensures $\Theta(\psi)=(2\pi)^{-1}\int_{0}^{2\pi}p(\theta-\psi)Z(\theta)d\theta=Z(\psi)$.
Given the periodic light pulse (Eq.~\eqref{eq:periodic_pulse}),
the entrainment phase $\psi_{st}$ (i.e. the circadian time at which
hamsters receive the light pulses) can be determined by 
\begin{equation}
\Omega-\omega+\chi\Theta(\psi_{st})=0,\hspace{1em}\Theta^{\prime}(\psi_{st})<0,\label{eq:st_cond}
\end{equation}
where we used Eq.~\eqref{eq:psi_dynamics}. Thus $\psi_{st}$ can
be given as a solution of $Z(\psi_{st})=(\omega-\Omega)/\chi$ with
$Z^{\prime}(\psi_{st})<0$. For the long ($\omega<\Omega$) and short
($\omega>\Omega$) pulses, $(\omega-\Omega)/\chi$ becomes negative
and positive, respectively. This shows that the long and short pulses
always act on hamsters at early ($\phi=1.3\sim3.2$; purple in Figs.~\ref{fig:optimal_pathway}(a)
and (b)) and late ($\phi=4.7\sim0.47$; orange in Figs.~\ref{fig:optimal_pathway}(a)
and (b)) subjective night, respectively. The effects of the light
pulse on the circadian clock depend on the concentration of $x_{k_{1}}$
and $x_{k_{2}}$ at these phases (Eq.~\eqref{eq:j_rate_equation}).
For the long pulse, we obtain $x_{\mathrm{LC},k_{1}}<x_{\mathrm{LC},k_{2}}$
(Fig.~\ref{fig:optimal_pathway}(b)), which indicates that the long
pulse always affects the expression of $x_{j_{2}}$ whereas it influences
$x_{j_{1}}$ only a little. In contrast, the short pulse affects $x_{j_{1}}$
more strongly than $x_{j_{2}}$. Our result shows that, provided the
circadian clocks are designed optimally, long and short pulses affect
the expression of two different components (\emph{Per}2 and \emph{Per}1)
differently. Surprisingly, our expression patterns agree with the
experiments of Schwartz \textit{et al.} \cite{Schwartz:2011:PerRoles}.
They hypothesized that light stimuli affect the transcription of \textit{Per}1
and the degradation of \textit{Per}2. In their molecular terms, $k_{1}$th
and $k_{2}$th species in our framework correspond to \emph{Bmal}1
and \textit{Per}2, which regulate the light effect (transcription
and degradation) on \textit{Per}1 and \textit{Per}2, respectively
(i.e. $x_{k_{1}}=[Bmal1]$ and $x_{k_{2}}=[Per2]$. Figure~\ref{fig:optimal_pathway}(c)).
The phase difference between \textit{Per}2 and \textit{Bmal}1 was
experimentally determined as $\nu\sim2$ \cite{Ueda:2002:MicroArrayTime}
and close to our result ($\nu=1.47$).

The pPRCs hitherto discussed are \emph{intrinsic} in the sense that
they represent the internal clock dynamics. The intrinsic pPRCs can
be observed only through the phase shift induced by short light pulses
\cite{Refinetti:2005:CircBook}. Theoretically, precise measurement
is possible only through $\delta$-peaked stimuli. In experiments
involving higher organisms, however, light pulses are much longer
than the $\delta$-peaked function, and observed pPRCs become different
from the intrinsic ones. To study the relation between intrinsic and
observed pPRCs, let us consider a squared-pulse stimulation $d\rho=\chi p(t)$
with $p(t)=\ell^{-1}H(t-t_{s})H(\ell+t_{s}-t)$ where $H(t)$ is the
Heaviside step function, $t_{s}$ is onset time of the pulses and
$\ell$ is the pulse duration. For $\ell\rightarrow0$, the squared
pulse reduces to a $\delta$-function $\delta(t-t_{s})$. Let $\tilde{Z}(\phi;\ell)$
be an observed pPRC of $Z(\phi)$ by a light pulse with the duration
$\ell$. Observed and intrinsic pPRCs can be related via $c_{\mu}=-\mathrm{i}\tilde{c}_{\mu}\ell\mu\Omega/\{\chi(1-\exp(\mathrm{i}\mu\Omega\ell))\}$
for $\mu\ne0$ and $c_{\mu}=\tilde{c}_{\mu}/\chi$ for $\mu=0$, where
$c_{\mu}$ and $\tilde{c}_{\mu}$ are Fourier coefficients of intrinsic
and observed pPRCs, respectively ($Z(\phi)=\sum_{\mu=-N_{\mu}}^{N_{\mu}}c_{\mu}\exp(\mathrm{i}\mu\phi)$
and $\tilde{Z}(\phi;\ell)=\sum_{\mu=-N_{\mu}}^{N_{\mu}}\tilde{c}_{\mu}\exp(\mathrm{i}\mu\phi)$
with $N_{\mu}$ being an expansion order). By this method, we inferred
the intrinsic pPRC from an observed pPRC in human \cite{Khalsa:2003:HumanPRC}
($\ell=6.7$-h) where a dead zone is seemingly nonexistent \cite{SI}. The inferred pPRC (dashed line) and the observed
pPRC (solid line) are shown in Fig.~\ref{fig:optimal_pathway}(d),
where the phase (horizontal axis) represents the onset of the pulse.
This result suggests that superficial pPRCs may lack a dead zone even
though their innate mechanisms actually do.

We have demonstrated that key properties of circadian clocks are consequences
of optimization to attain the maximal limit of entrainability and
regularity. Our theory explains known experimental results such as
the role sharing of two inputs and different gene expression patterns
by different pulses. We also explain the superficial absence of a
dead zone in human. The model can be used to reveal key molecular
elements responsible for the clock.

This work was supported by Grant-in-Aid for Young Scientists B (Y.H.:
No.~25870171) and for Innovative Areas ``Biosynthetic machinery''
(M.A.) from MEXT, Japan.


\begin{thebibliography}{22}
\expandafter\ifx\csname natexlab\endcsname\relax\def\natexlab#1{#1}\fi
\expandafter\ifx\csname bibnamefont\endcsname\relax
  \def\bibnamefont#1{#1}\fi
\expandafter\ifx\csname bibfnamefont\endcsname\relax
  \def\bibfnamefont#1{#1}\fi
\expandafter\ifx\csname citenamefont\endcsname\relax
  \def\citenamefont#1{#1}\fi
\expandafter\ifx\csname url\endcsname\relax
  \def\url#1{\texttt{#1}}\fi
\expandafter\ifx\csname urlprefix\endcsname\relax\def\urlprefix{URL }\fi
\providecommand{\bibinfo}[2]{#2}
\providecommand{\eprint}[2][]{\url{#2}}

\bibitem[{\citenamefont{Young and Kay}(2001)}]{Young:2001:GeneticsCirc}
\bibinfo{author}{\bibfnamefont{M.~W.} \bibnamefont{Young}} \bibnamefont{and}
  \bibinfo{author}{\bibfnamefont{S.~A.} \bibnamefont{Kay}},
  \bibinfo{journal}{Nat. Rev. Genet.} \textbf{\bibinfo{volume}{2}},
  \bibinfo{pages}{702} (\bibinfo{year}{2001}).

\bibitem[{\citenamefont{Johnson et~al.}(2011)\citenamefont{Johnson, Stewart,
  and Egli}}]{Johnson:2011:CyanoCircadian}
\bibinfo{author}{\bibfnamefont{C.~H.} \bibnamefont{Johnson}},
  \bibinfo{author}{\bibfnamefont{P.~L.} \bibnamefont{Stewart}},
  \bibnamefont{and} \bibinfo{author}{\bibfnamefont{M.}~\bibnamefont{Egli}},
  \bibinfo{journal}{Annu. Rev. Biophys.} \textbf{\bibinfo{volume}{40}},
  \bibinfo{pages}{143} (\bibinfo{year}{2011}).

\bibitem[{\citenamefont{Vilar et~al.}(2002)\citenamefont{Vilar, Kueh, Barkai,
  and Leibler}}]{Vilar:2002:NoiseResistGeneOsc}
\bibinfo{author}{\bibfnamefont{J.~M.~G.} \bibnamefont{Vilar}},
  \bibinfo{author}{\bibfnamefont{H.~Y.} \bibnamefont{Kueh}},
  \bibinfo{author}{\bibfnamefont{N.}~\bibnamefont{Barkai}}, \bibnamefont{and}
  \bibinfo{author}{\bibfnamefont{S.}~\bibnamefont{Leibler}},
  \bibinfo{journal}{PNAS} \textbf{\bibinfo{volume}{99}}, \bibinfo{pages}{5988}
  (\bibinfo{year}{2002}).

\bibitem[{\citenamefont{Gonze et~al.}(2002)\citenamefont{Gonze, Halloy, and
  Goldbeter}}]{Gonze:2002:RobustCircadian}
\bibinfo{author}{\bibfnamefont{D.}~\bibnamefont{Gonze}},
  \bibinfo{author}{\bibfnamefont{J.}~\bibnamefont{Halloy}}, \bibnamefont{and}
  \bibinfo{author}{\bibfnamefont{A.}~\bibnamefont{Goldbeter}},
  \bibinfo{journal}{PNAS} \textbf{\bibinfo{volume}{99}}, \bibinfo{pages}{673}
  (\bibinfo{year}{2002}).

\bibitem[{\citenamefont{Herzog et~al.}(2004)\citenamefont{Herzog, Aton, Numano,
  Sakaki, and Tei}}]{Herzon:2004:CircadianPrecision}
\bibinfo{author}{\bibfnamefont{E.~D.} \bibnamefont{Herzog}},
  \bibinfo{author}{\bibfnamefont{S.~J.} \bibnamefont{Aton}},
  \bibinfo{author}{\bibfnamefont{R.}~\bibnamefont{Numano}},
  \bibinfo{author}{\bibfnamefont{Y.}~\bibnamefont{Sakaki}}, \bibnamefont{and}
  \bibinfo{author}{\bibfnamefont{H.}~\bibnamefont{Tei}}, \bibinfo{journal}{J.
  Biol. Rhythms} \textbf{\bibinfo{volume}{19}}, \bibinfo{pages}{35}
  (\bibinfo{year}{2004}).

\bibitem[{\citenamefont{Gonze and Goldbeter}(2000)}]{Gonze:2000:Entrainment}
\bibinfo{author}{\bibfnamefont{D.}~\bibnamefont{Gonze}} \bibnamefont{and}
  \bibinfo{author}{\bibfnamefont{A.}~\bibnamefont{Goldbeter}},
  \bibinfo{journal}{J. Stat. Phys.} \textbf{\bibinfo{volume}{101}},
  \bibinfo{pages}{649} (\bibinfo{year}{2000}).

\bibitem[{\citenamefont{Roenneberg et~al.}(2003)\citenamefont{Roenneberg, Daan,
  and Merrow}}]{Roenneberg:2003:ArtEntrain}
\bibinfo{author}{\bibfnamefont{T.}~\bibnamefont{Roenneberg}},
  \bibinfo{author}{\bibfnamefont{S.}~\bibnamefont{Daan}}, \bibnamefont{and}
  \bibinfo{author}{\bibfnamefont{M.}~\bibnamefont{Merrow}},
  \bibinfo{journal}{J. Biol. Rhythms} \textbf{\bibinfo{volume}{18}},
  \bibinfo{pages}{183} (\bibinfo{year}{2003}).

\bibitem[{\citenamefont{Golombek and
  Rosenstein}(2010)}]{Golombek:2010:CircadianEntrainment}
\bibinfo{author}{\bibfnamefont{D.~A.} \bibnamefont{Golombek}} \bibnamefont{and}
  \bibinfo{author}{\bibfnamefont{R.~E.} \bibnamefont{Rosenstein}},
  \bibinfo{journal}{Physiol. Rev.} \textbf{\bibinfo{volume}{90}},
  \bibinfo{pages}{1063} (\bibinfo{year}{2010}).

\bibitem[{\citenamefont{Hasegawa and Arita}(2014)}]{Hasegawa:2013:OptimalPRC}
\bibinfo{author}{\bibfnamefont{Y.}~\bibnamefont{Hasegawa}} \bibnamefont{and}
  \bibinfo{author}{\bibfnamefont{M.}~\bibnamefont{Arita}}, \bibinfo{journal}{J.
  R. Soc. Interface} \textbf{\bibinfo{volume}{11}}, \bibinfo{pages}{20131018}
  (\bibinfo{year}{2014}).

\bibitem[{\citenamefont{Troein et~al.}(2009)\citenamefont{Troein, Locke,
  Turner, and Millar}}]{Troein:2009:ComplexClocks}
\bibinfo{author}{\bibfnamefont{C.}~\bibnamefont{Troein}},
  \bibinfo{author}{\bibfnamefont{J.~C.} \bibnamefont{Locke}},
  \bibinfo{author}{\bibfnamefont{M.~S.} \bibnamefont{Turner}},
  \bibnamefont{and} \bibinfo{author}{\bibfnamefont{A.~J.}
  \bibnamefont{Millar}}, \bibinfo{journal}{Curr. Biol.}
  \textbf{\bibinfo{volume}{19}}, \bibinfo{pages}{1961} (\bibinfo{year}{2009}).

\bibitem[{\citenamefont{Troein et~al.}(2011)\citenamefont{Troein, Corellou,
  Dixon, van Ooijen, O'Neill, Bouget, and Millar}}]{Troein:2011:TauriClock}
\bibinfo{author}{\bibfnamefont{C.}~\bibnamefont{Troein}},
  \bibinfo{author}{\bibfnamefont{F.}~\bibnamefont{Corellou}},
  \bibinfo{author}{\bibfnamefont{L.~E.} \bibnamefont{Dixon}},
  \bibinfo{author}{\bibfnamefont{G.}~\bibnamefont{van Ooijen}},
  \bibinfo{author}{\bibfnamefont{J.~S.} \bibnamefont{O'Neill}},
  \bibinfo{author}{\bibfnamefont{F.-Y.} \bibnamefont{Bouget}},
  \bibnamefont{and} \bibinfo{author}{\bibfnamefont{A.~J.}
  \bibnamefont{Millar}}, \bibinfo{journal}{Plant J.}
  \textbf{\bibinfo{volume}{66}}, \bibinfo{pages}{375} (\bibinfo{year}{2011}).

\bibitem[{\citenamefont{Albrecht et~al.}(2001)\citenamefont{Albrecht, Zheng,
  Larkin, Sun, and Lee}}]{Albrecht:2001:mPer1_mPer2}
\bibinfo{author}{\bibfnamefont{U.}~\bibnamefont{Albrecht}},
  \bibinfo{author}{\bibfnamefont{B.}~\bibnamefont{Zheng}},
  \bibinfo{author}{\bibfnamefont{D.}~\bibnamefont{Larkin}},
  \bibinfo{author}{\bibfnamefont{Z.~S.} \bibnamefont{Sun}}, \bibnamefont{and}
  \bibinfo{author}{\bibfnamefont{C.~C.} \bibnamefont{Lee}},
  \bibinfo{journal}{J. Biol. Rhythms} \textbf{\bibinfo{volume}{16}},
  \bibinfo{pages}{100} (\bibinfo{year}{2001}).

\bibitem[{\citenamefont{Steinlechner et~al.}(2002)\citenamefont{Steinlechner,
  Jacobmeier, Scherbarth, Dernbach, Kruse, and
  Albrecht}}]{Steinlechner:2002:MutantLL}
\bibinfo{author}{\bibfnamefont{S.}~\bibnamefont{Steinlechner}},
  \bibinfo{author}{\bibfnamefont{B.}~\bibnamefont{Jacobmeier}},
  \bibinfo{author}{\bibfnamefont{F.}~\bibnamefont{Scherbarth}},
  \bibinfo{author}{\bibfnamefont{H.}~\bibnamefont{Dernbach}},
  \bibinfo{author}{\bibfnamefont{F.}~\bibnamefont{Kruse}}, \bibnamefont{and}
  \bibinfo{author}{\bibfnamefont{U.}~\bibnamefont{Albrecht}},
  \bibinfo{journal}{J. Biol. Rhythms} \textbf{\bibinfo{volume}{17}},
  \bibinfo{pages}{202} (\bibinfo{year}{2002}).

\bibitem[{\citenamefont{Refinetti}(2005)}]{Refinetti:2005:CircBook}
\bibinfo{author}{\bibfnamefont{R.}~\bibnamefont{Refinetti}},
  \emph{\bibinfo{title}{Circadian physiology}} (\bibinfo{publisher}{Taylor \&
  Francis}, \bibinfo{year}{2005}), \bibinfo{edition}{2nd} ed.

\bibitem[{\citenamefont{Schwartz et~al.}(2011)\citenamefont{Schwartz,
  Tavakoli-Nezhad, Lambert, Weaver, and de~la
  Iglesia}}]{Schwartz:2011:PerRoles}
\bibinfo{author}{\bibfnamefont{W.~J.} \bibnamefont{Schwartz}},
  \bibinfo{author}{\bibfnamefont{M.}~\bibnamefont{Tavakoli-Nezhad}},
  \bibinfo{author}{\bibfnamefont{C.~M.} \bibnamefont{Lambert}},
  \bibinfo{author}{\bibfnamefont{D.~R.} \bibnamefont{Weaver}},
  \bibnamefont{and} \bibinfo{author}{\bibfnamefont{H.~O.} \bibnamefont{de~la
  Iglesia}}, \bibinfo{journal}{PNAS} \textbf{\bibinfo{volume}{108}},
  \bibinfo{pages}{17219} (\bibinfo{year}{2011}).

\bibitem[{\citenamefont{Kuramoto}(2003)}]{Kuramoto:2003:OscBook}
\bibinfo{author}{\bibfnamefont{Y.}~\bibnamefont{Kuramoto}},
  \emph{\bibinfo{title}{Chemical Oscillations, Waves, and Turbulence}}
  (\bibinfo{publisher}{Dover publications}, \bibinfo{address}{Mineola, New
  York}, \bibinfo{year}{2003}).

\bibitem[{\citenamefont{Taylor et~al.}(2008)\citenamefont{Taylor, Gunawan,
  Petzold, and {Doyle III}}}]{Taylor:2008:Sensitivity}
\bibinfo{author}{\bibfnamefont{S.~R.} \bibnamefont{Taylor}},
  \bibinfo{author}{\bibfnamefont{R.}~\bibnamefont{Gunawan}},
  \bibinfo{author}{\bibfnamefont{L.~R.} \bibnamefont{Petzold}},
  \bibnamefont{and} \bibinfo{author}{\bibfnamefont{F.~J.} \bibnamefont{{Doyle
  III}}}, \bibinfo{journal}{IEEE Trans. Automat. Contr.}
  \textbf{\bibinfo{volume}{53}}, \bibinfo{pages}{177} (\bibinfo{year}{2008}).



\bibitem[{\citenamefont{Hasegawa and Arita}(2013)}]{Hasegawa:2012:GeneOsc}
\bibinfo{author}{\bibfnamefont{Y.}~\bibnamefont{Hasegawa}} \bibnamefont{and}
  \bibinfo{author}{\bibfnamefont{M.}~\bibnamefont{Arita}}, \bibinfo{journal}{J.
  R. Soc. Interface} \textbf{\bibinfo{volume}{10}}, \bibinfo{pages}{20121020}
  (\bibinfo{year}{2013}).

\bibitem{SI}
  See Supplemental Material, which includes Ref.~\cite{Storn:1997:DE}.

\bibitem[{\citenamefont{Storn and Price}(1997)}]{Storn:1997:DE}
\bibinfo{author}{\bibfnamefont{R.}~\bibnamefont{Storn}} \bibnamefont{and}
  \bibinfo{author}{\bibfnamefont{K.}~\bibnamefont{Price}}, \bibinfo{journal}{J.
  Glob. Optim.} \textbf{\bibinfo{volume}{11}}, \bibinfo{pages}{341}
  (\bibinfo{year}{1997}).

\bibitem[{\citenamefont{Johnsson and
  Engelmann}(2007)}]{Johnsson:2007:CircLightReview}
\bibinfo{author}{\bibfnamefont{A.}~\bibnamefont{Johnsson}} \bibnamefont{and}
  \bibinfo{author}{\bibfnamefont{W.}~\bibnamefont{Engelmann}}, in
  \emph{\bibinfo{booktitle}{Photobiology: The Science of Life and Light}}
  (\bibinfo{publisher}{Springer}, \bibinfo{year}{2007}).

\bibitem[{\citenamefont{Hartmann}(1994)}]{Hartmann:1994:PhysClimate}
\bibinfo{author}{\bibfnamefont{D.~L.} \bibnamefont{Hartmann}},
  \emph{\bibinfo{title}{Global Physical Climatology}}
  (\bibinfo{publisher}{Academic Press}, \bibinfo{year}{1994}).

\bibitem[{\citenamefont{Khalsa et~al.}(2003)\citenamefont{Khalsa, Jewett,
  Cajochen, and Czeisler}}]{Khalsa:2003:HumanPRC}
\bibinfo{author}{\bibfnamefont{S.~B.~S.} \bibnamefont{Khalsa}},
  \bibinfo{author}{\bibfnamefont{M.~E.} \bibnamefont{Jewett}},
  \bibinfo{author}{\bibfnamefont{C.}~\bibnamefont{Cajochen}}, \bibnamefont{and}
  \bibinfo{author}{\bibfnamefont{C.~A.} \bibnamefont{Czeisler}},
  \bibinfo{journal}{J. Physiol.} \textbf{\bibinfo{volume}{549}},
  \bibinfo{pages}{945} (\bibinfo{year}{2003}).

\bibitem[{\citenamefont{Ueda et~al.}(2002)\citenamefont{Ueda, Chen, Adachi,
  Wakamatsu, Hayashi, Takasugi, Nagano, Nakahama, Suzuki, Sugano
  et~al.}}]{Ueda:2002:MicroArrayTime}
\bibinfo{author}{\bibfnamefont{H.~R.} \bibnamefont{Ueda}},
  \bibinfo{author}{\bibfnamefont{W.}~\bibnamefont{Chen}},
  \bibinfo{author}{\bibfnamefont{A.}~\bibnamefont{Adachi}},
  \bibinfo{author}{\bibfnamefont{H.}~\bibnamefont{Wakamatsu}},
  \bibinfo{author}{\bibfnamefont{S.}~\bibnamefont{Hayashi}},
  \bibinfo{author}{\bibfnamefont{T.}~\bibnamefont{Takasugi}},
  \bibinfo{author}{\bibfnamefont{M.}~\bibnamefont{Nagano}},
  \bibinfo{author}{\bibfnamefont{K.-i.} \bibnamefont{Nakahama}},
  \bibinfo{author}{\bibfnamefont{Y.}~\bibnamefont{Suzuki}},
  \bibinfo{author}{\bibfnamefont{S.}~\bibnamefont{Sugano}},
  \bibnamefont{et~al.}, \bibinfo{journal}{Nature}
  \textbf{\bibinfo{volume}{418}}, \bibinfo{pages}{534} (\bibinfo{year}{2002}).

\end{thebibliography}
\end{document}


\title{Supplemental Material for\\
``Optimal Implementations for Reliable Circadian Clocks''}

\author{Yoshihiko Hasegawa and Masanori Arita}

\maketitle
This supplemental material describes detailed calculations introduced
in the main text. Equation and figure numbers in this section are
prefixed with S (e.g., Eq.\textbf{~}(S1) or Fig.~S1). Numbers without
the prefix (e.g., Eq.~(1) or Fig.~1) refer to numbers in the main
text.

\section{Optimization}

We calculated the optimal phase-response curves (PRCs) using two approaches,
variational and numerical methods, whose explicit procedures are described
below.

\subsection{Variational method}

Regularity $\mathcal{V}_{T}$ and entrainability $\mathcal{E}$ are
calculated in our previous study~\cite{Hasegawa:2013:OptimalPRC}.
Regularity is defined by the period variance of the oscillation (higher
regularity corresponds to smaller period variance), which is represented
by
\begin{equation}
\mathcal{V}_{T}=\frac{T^{3}}{4\pi^{3}}\int_{0}^{2\pi}\sum_{i=1}^{N}U_{i}(\theta)^{2}Q_{i}(\theta)^{2}d\theta.\label{eq:def_varT}
\end{equation}
Here $N$ is the dimension, $U_{i}(\phi)$ is the infinitesimal PRC (iPRC), $T$ is the period
of the unperturbed oscillator and $Q_{i}(\phi)$ is a multiplicative
noise term (see the main text). Entrainability, the extent of synchronization
to a periodic input signal, is defined by the width of the Arnold
tongue. Given a periodic input signal $p(\omega t)$ ($\omega$ is
angular frequency), entrainability is represented by
\begin{align}
\mathcal{E}&=\Theta(\psi_{M})-\Theta(\psi_{m}),\label{eq:E_line1}\\
&=\frac{1}{2\pi}\int_{0}^{2\pi}Z(\theta)\left\{ p(\theta-\psi_{M})-p(\theta-\psi_{m})\right\} d\theta,\label{eq:def_ent}
\end{align}
where $\Theta(\psi)=(2\pi)^{-1}\int_{0}^{2\pi}Z(\psi+\theta)p(\theta)d\theta$,
$\psi_{M}=\mathrm{argmax}_{\psi}\Theta(\psi)$, $\psi_{m}=\mathrm{argmin}_{\psi}\Theta(\psi)$ and 
$Z(\phi)$ is the parametric PRC (pPRC) defined in Eq.~\PRCUZ.

The optimal PRCs, which maximize entrainability $\mathcal{E}$ under
constant regularity $\mathcal{V}_{T}=\sigma_{T}^{2}$, can be calculated
by the Euler--Lagrange variational method. A variational equation
to be optimized is
\begin{equation}
\mathcal{L}=\mathcal{E}-\lambda\mathcal{V}_{T},\label{eq:variational_equation}
\end{equation}
where $\lambda$ is a Lagrange multiplier. With Eqs.~\PRCUZ, \eqref{eq:def_varT} and \eqref{eq:def_ent}, 
Eq.~\eqref{eq:variational_equation} is specifically given by
\begin{align}
\mathcal{L} & =\frac{1}{2\pi}\int_{0}^{2\pi}\sum_{i=1}^{N}\left\{ \left\{ p(\theta-\psi_{M})-p(\theta-\psi_{m})\right\} U_{i}(\theta)\frac{\partial F_{i}(\theta;\rho)}{\partial\rho}-\frac{\lambda T^{3}}{2\pi^{2}}U_{i}(\theta)^{2}Q_{i}(\theta)^{2}\right\} d\theta,\label{eq:variational_equation2}
\end{align}
where $F_{i}(\phi;\rho)$ is the $i$th reaction rate and $\rho$ is a light-sensitive parameter.
The functional derivative of $\mathcal{L}$ with respect to $U_{i}$
yields
\begin{align}
\frac{\delta\mathcal{L}}{\delta U_{i}} & =\left\{ p(\theta-\psi_{M})-p(\theta-\psi_{m})\right\} \frac{\partial F_{i}(\theta;\rho)}{\partial\rho}-\frac{\lambda T^{3}}{\pi^{2}}U_{i}(\theta)Q_{i}(\theta)^{2},\nonumber \\
 & =0.\label{eq:stationary_condition}
\end{align}
Solving Eq.~\eqref{eq:stationary_condition}, we obtain the optimal
PRCs $U_{i}(\phi)$ and $Z(\phi)$ given by
\begin{align}
U_{i}(\phi) & =\frac{\pi^{2}}{T^{3}\lambda}\frac{p(\phi-\psi_{M})-p(\phi-\psi_{m})}{Q_{i}(\phi)^{2}}\frac{\partial F_{i}(\phi;\rho)}{\partial\rho},\label{eq:Ui_def}\\
Z(\phi) & =\frac{\pi^{2}}{T^{3}\lambda}\sum_{i=1}^{N}\frac{p(\phi-\psi_{M})-p(\phi-\psi_{m})}{Q_{i}(\phi)^{2}}\left(\frac{\partial F_{i}(\phi;\rho)}{\partial\rho}\right)^{2},\label{eq:Z_def}
\end{align}
where the Lagrange multiplier $\lambda$ is 
\begin{equation}
\lambda=\sqrt{\frac{\pi}{4T^{3}\sigma_{T}^{2}}\int_{0}^{2\pi}\sum_{i=1}^{N}\frac{[p(\theta-\psi_{M})-p(\theta-\psi_{m})]^{2}}{Q_{i}(\theta)^{2}}\left(\frac{\partial F_{i}(\theta;\rho)}{\partial\rho}\right)^{2}d\theta}.\label{eq:Lagrange_def}
\end{equation}
A substitution of Eq.~\eqref{eq:Z_def} into Eq.~\eqref{eq:def_ent}
yields the entrainability $\mathcal{E}$ represented by
\begin{align}
\mathcal{E} & =\frac{1}{2\pi}\int_{0}^{2\pi}\left\{ p(\theta-\psi_{M})-p(\theta-\psi_{m})\right\} \sum_{i=1}^{N}U_{i}(\theta)\frac{\partial F_{i}(\theta;\rho)}{\partial\rho}d\theta,\nonumber \\
 & =\frac{\pi}{2T^{3}\lambda}\int_{0}^{2\pi}\sum_{i=1}^{N}\frac{\left\{ p(\theta-\psi_{M})-p(\theta-\psi_{m})\right\} ^{2}}{Q_{i}(\theta)^{2}}\left(\frac{\partial F_{i}(\theta;\rho)}{\partial\rho}\right)^{2}d\theta.\label{eq:entrainability}
\end{align}
We maximize $\mathcal{E}$ as functions of $\Delta$ and $\delta$,
where $\Delta=\psi_{M}-\psi_{m}$ and $\delta=\psi_{m}$. Substituting
the obtained parameters into Eqs.~\eqref{eq:Ui_def} and \eqref{eq:Z_def},
we can calculate the optimal PRCs \cite{Hasegawa:2013:OptimalPRC}. 

The dead zone length $L$ is defined by the length of null parts in
the PRCs. Because a dead zone emerges when $\Delta\ne\pi$ in Eqs.~\eqref{eq:Ui_def}
and \eqref{eq:Z_def}, we can naturally define its length by
\begin{equation}
L=|\Delta-\pi|,\label{eq:DZL}
\end{equation}
which is plotted in Fig.~\FIGdiffUdependence(b) as a function of
the phase difference $\nu$.

\subsection{Numerical method}

\begin{figure}
\begin{centering}
\includegraphics[width=8cm]{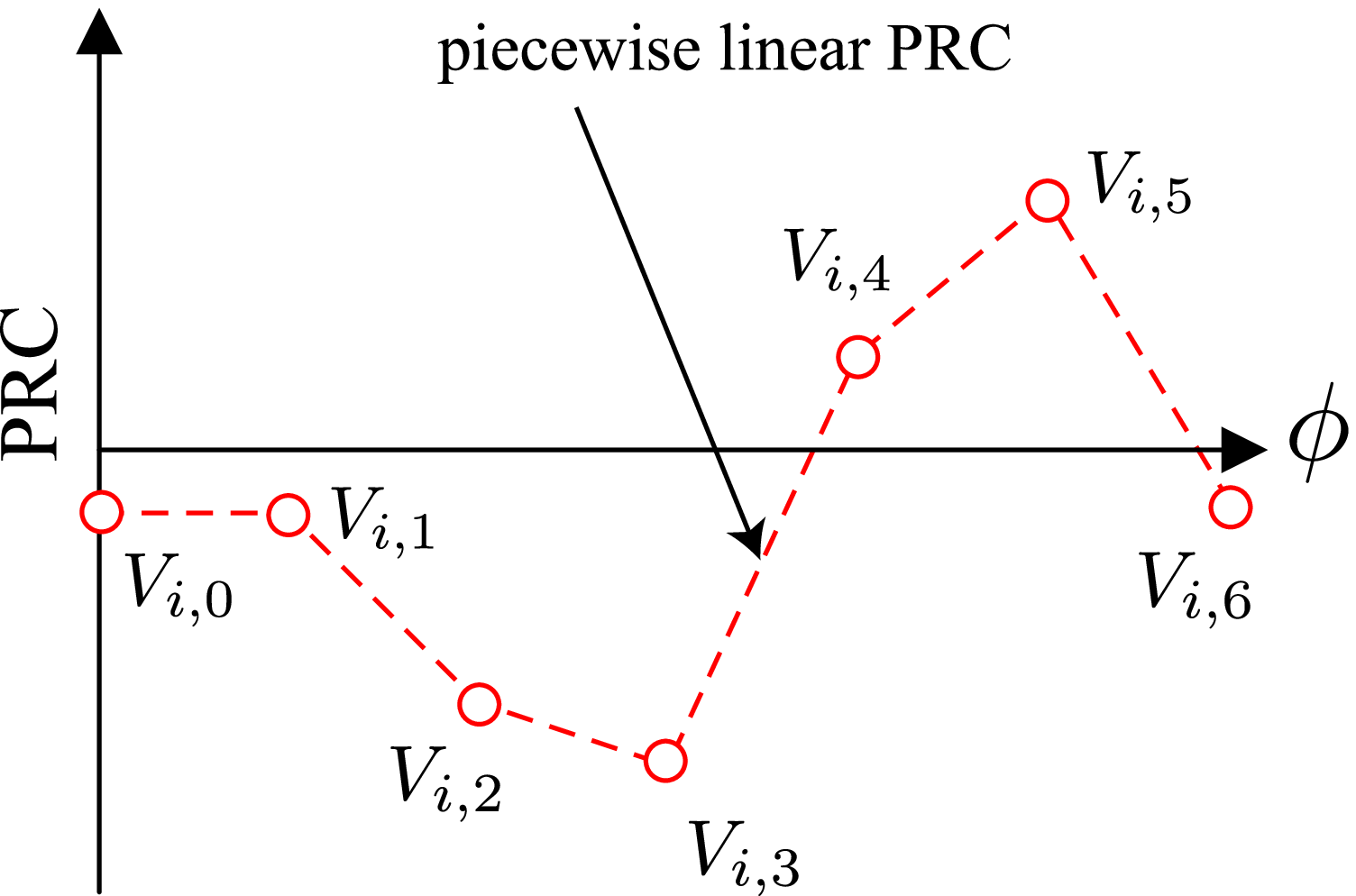}
\par\end{centering}

\caption{Piecewise linear PRC ($K=6$) as an approximation of the iPRC $U_{i}(\phi)$,
where the $l$th knot is $V_{i,l}$ (circle). We set $V_{i,K}=V_{i,0}$
for a periodic condition. \label{fig:PWL}}

\end{figure}

Because the optimal iPRC $U_{i}(\phi)$ and parameters ($\psi_{M}$
and $\psi_{m}$) are interdependent, the variational method is not
trivial. Consequently, to verify the correctness of the variational
method numerically, we calculated the optimal PRCs with the evolutionary
algorithm (specifically, we used a differential evolution (DE) of
Storn and Price \cite{Storn:1997:DE} provided by \textit{MATHEMATICA}
9). Dividing the iPRC $U_{i}(\phi)$ into a piecewise linear function,
we can reduce the variational problem to a conventional multivariate
parameter optimization problem. We divide the iPRC $U_{i}(\phi)$
into $K$ regions (Fig.~\ref{fig:PWL}) and linearly connect each
knot to create a piecewise linear function as an approximation to
$U_{i}$, where the value of the $l$th knot is given by $V_{i,l}$
for $l=0,1,..,K-1$ (we set $V_{i,K}=V_{i,0}$ for a periodic condition):
\[
U_{i}(\phi)\simeq\mathrm{LF}(\phi;\boldsymbol{V}_{i}).
\]
Here $\mathrm{LF}(\phi;\boldsymbol{V}_{i})$ is a piecewise linear
function (LF is short for linear function) whose knots are $\boldsymbol{V}_{i}=(V_{i,0},V_{i,1},..,V_{i,K-1})$.
Based on Eq.~\jUrateUequation~in the main text, we let $U_{j_{i}}(\phi)$
be the iPRC of molecular species that have an $i$th light entry point
(for the $M$ input case there are $M$ iPRCs $U_{j_{1}}(\phi),..,U_{j_{M}}(\phi)$).
The constraint on the period variance is taken into account by considering
scaled knots $\kappa\boldsymbol{V}_{i}=(\kappa V_{i,0},..,\kappa V_{i,K-1})$
where $\kappa$ is a scaling parameter determined by the variance
constraint ($\kappa$ is determined so that a piecewise linear PRC
with $\kappa V_{i,0},\kappa V_{i,1},..,\kappa V_{i,K-1}$ yields the
period variance $\mathcal{V}_{T}=\sigma_{T}^{2}=1$). The period variance
(Eq.~\eqref{eq:def_varT}) for $\mathrm{LF}(\phi;\kappa\boldsymbol{V}_{j_{i}})$
is given by
\begin{equation}
\sigma_{T}^{2}=\frac{T^{3}}{4\pi^{3}}\int_{0}^{2\pi}\sum_{i=1}^{M}\mathrm{LF}(\theta;\kappa\boldsymbol{V}_{j_{i}})^{2}Q_{j_{i}}(\theta)^{2}d\theta=\frac{\kappa^{2}T^{3}}{4\pi^{3}}\int_{0}^{2\pi}\sum_{i=1}^{M}\mathrm{LF}(\theta;\boldsymbol{V}_{j_{i}})^{2}Q_{j_{i}}(\theta)^{2}d\theta.\label{eq:kappa_constraint}
\end{equation}
We adopted the assumption that the noise term $Q_{i}(\phi)$ is only
present in input molecules $i=j_{1},j_{2},..,j_{M}$. Therefore, 
$\kappa$ yielding the desired period variance $\sigma_{T}^{2}$
is calculated as
\begin{equation}
\kappa=\sqrt{\frac{4\pi^{3}\sigma_{T}^{2}}{T^{3}\int_{0}^{2\pi}\sum_{i=1}^{M}\mathrm{LF}(\theta;\boldsymbol{V}_{j_{i}})^{2}Q_{j_{i}}(\theta)^{2}d\theta}}.\label{eq:kappa_def}
\end{equation}
Because $\mathrm{LF}(\phi;\boldsymbol{V}_{i})$ is a linear function,
the integral in Eq.~\eqref{eq:kappa_def} can be calculated in a
closed form. 

We can calculate the entrainability $\mathcal{E}$ with
respect to the piecewise linear functions. We first calculate $\Theta(\psi)$ for $\mathrm{LF}(\phi;\boldsymbol{V}_i)$ (again the integral can be calculated in a closed form):
\begin{align}
\Theta(\psi;\boldsymbol{V}_{j_{1}},..,\boldsymbol{V}_{j_{M}}) & =\frac{1}{2\pi}\int_{0}^{2\pi}p(\theta-\psi)\sum_{i=1}^{M}\mathrm{LF}(\theta;\kappa\boldsymbol{V}_{j_{i}})\frac{\partial F_{j_{i}}(\theta;\rho)}{\partial\rho}d\theta,\nonumber \\
 & =\frac{\kappa}{2\pi}\int_{0}^{2\pi}p(\theta-\psi)\sum_{i=1}^{M}\mathrm{LF}(\theta;\boldsymbol{V}_{j_{i}})\frac{\partial F_{j_{i}}(\theta;\rho)}{\partial\rho}d\theta.\label{eq:Theta_do_def}
\end{align}
From Eq.~\eqref{eq:E_line1}, the entrainability is given by
\[
\mathcal{E}(\Delta,\delta,\boldsymbol{V}_{j_{1}},..,\boldsymbol{V}_{j_{M}})=\Theta(\Delta+\delta;\boldsymbol{V}_{j_{1}},..,\boldsymbol{V}_{j_{M}})-\Theta(\delta;\boldsymbol{V}_{j_{1}},..,\boldsymbol{V}_{j_{M}}),
\]
which is a function of $\Delta=\psi_{M}-\psi_{m}$, $\delta=\psi_{m}$
and $\boldsymbol{V}_{j_{1}},..,\boldsymbol{V}_{j_{M}}$ ($\boldsymbol{V}_{i}\in\mathbb{R}^{K}$).
For $M$ iPRCs ($M$ input pathways), $M\times K+2$ parameters have
to be optimized. The pPRC is calculated
by Eq.~\PRCUZ:
\begin{align}
Z(\phi) & \simeq\sum_{i=1}^{M}\mathrm{LF}(\phi;\kappa\boldsymbol{V}_{j_{i}})\frac{\partial F_{j_{i}}(\phi;\rho)}{\partial\rho},\label{eq:Z_LF}\\
Z_{j_{i}}(\phi) & \simeq\mathrm{LF}(\phi;\kappa\boldsymbol{V}_{j_{i}})\frac{\partial F_{j_{i}}(\phi;\rho)}{\partial\rho}.\label{eq:Z_LF2}
\end{align}
$Z_{j_{1}}$and $Z_{j_{2}}$ by the evolutionary algorithm with $K=20$
are plotted in Figs.~\FIGoptimalUPRCUtwo(a) and (b) with circles
and triangles, respectively (dashed and dot-dashed lines reflect the
variational method).

\section{Period dependence on constant light intensity}

Under a constant light condition, the period may vary, depending on
the light intensity. When a parameter $\rho$ is perturbed $\rho+d\rho$
with $d\rho=\chi p(t)$ (for the constant light condition, $p(t)=1$),
the phase $\phi\in[0,2\pi)$ obeys the following differential equation
\begin{equation}
\frac{d\phi}{dt}=\Omega+\chi Z(\phi)p(t),\label{eq:dphidt}
\end{equation}
where $\Omega=2\pi/T$ is angular frequency of the oscillator and
$\chi$ is the signal strength. In order to discuss a solution, we
employ a perturbation expansion method. For the first order expansion,
we assume that the solution $\phi(t)$ can be represented by 
\begin{equation}
\phi(t)=\phi_{0}(t)+\chi\phi_{1}(t),\label{eq:PE}
\end{equation}
where $\phi_{0}(t)$ and $\phi_{1}(t)$ are zeroth- and first-order
terms, respectively. Substituting Eq.~\eqref{eq:PE} into Eq.~\eqref{eq:dphidt}
to obtain
\begin{align}
\frac{d\phi_{0}}{dt}+\chi\frac{d\phi_{1}}{dt} & =\Omega+\chi Z(\phi_{0}+\chi\phi_{1})p(t),\nonumber \\
 & \simeq\Omega+\chi\left\{ Z(\phi_{0})+\chi\phi_{1}Z^{\prime}(\phi_{0})\right\} p(t),\label{eq:de2}
\end{align}
where we used $Z(\phi_{0}+\chi\phi_{1})\simeq Z(\phi_{0})+\chi\phi_{1}Z^{\prime}(\phi_{0})$.
Equating Eq.~\eqref{eq:de2} with respect to the order of $\chi$,
we have the following coupled differential equations:
\begin{align}
O(1)\hspace{1em} & \frac{d\phi_{0}}{dt}=\Omega,\label{eq:zeroth_order}\\
O(\chi)\hspace{1em} & \frac{d\phi_{1}}{dt}=Z(\phi_{0})p(t).\label{eq:first_order}
\end{align}
Integrating Eqs.~\eqref{eq:zeroth_order} and \eqref{eq:first_order}
from $t=0$ to $t=t_{e}$, where $t_{e}$ is sufficiently large time,
with an initial condition $\phi(0)=0$, we have the following formal
solutions:
\begin{align}
\phi_{0}(t) & =\Omega t,\label{eq:phi0}\\
\phi_{1}(t_{e}) & =\int_{0}^{t_{e}}Z(\Omega t)p(t)dt.\label{eq:phi1}
\end{align}
From Eqs.~\eqref{eq:PE}, \eqref{eq:phi0} and \eqref{eq:phi1},
the phase $\phi$ at time $t_{e}=T$ with the input strength $\chi$
is given by
\begin{align}
\phi(T;\chi) & =\phi_{0}(T)+\chi\phi_{1}(T),\nonumber \\
 & =2\pi+\frac{\chi T}{2\pi}\int_{0}^{2\pi}Z(\theta)d\theta,\label{eq:phi_T_chi}
\end{align}
where we used $p(t)=1$. For weak $\chi$, the period $T_{\chi}$,
which is the period under the constant light condition with intensity
$\chi$, is approximated by
\begin{equation}
\frac{T_{\chi}}{T}\simeq\frac{\phi(T;0)}{\phi(T;\chi)}=\frac{2\pi}{\phi(T;\chi)}.\label{eq:Tchi_T}
\end{equation}
Substituting Eq.~\eqref{eq:phi_T_chi} into Eq.~\eqref{eq:Tchi_T},
we have
\begin{align}
\frac{T_{\chi}}{T} & \simeq\frac{1}{{\displaystyle 1+\frac{\chi T}{4\pi^{2}}\int_{0}^{2\pi}Z(\theta)d\theta}},\nonumber \\
 & \simeq1-\frac{\chi T}{4\pi^{2}}\int_{0}^{2\pi}Z(\theta)d\theta,\label{eq:T_chi_T_2}
\end{align}
which is an equation shown in the main text.

\section{Inference of intrinsic pPRC}

\begin{figure}
\begin{centering}
\includegraphics[width=13cm]{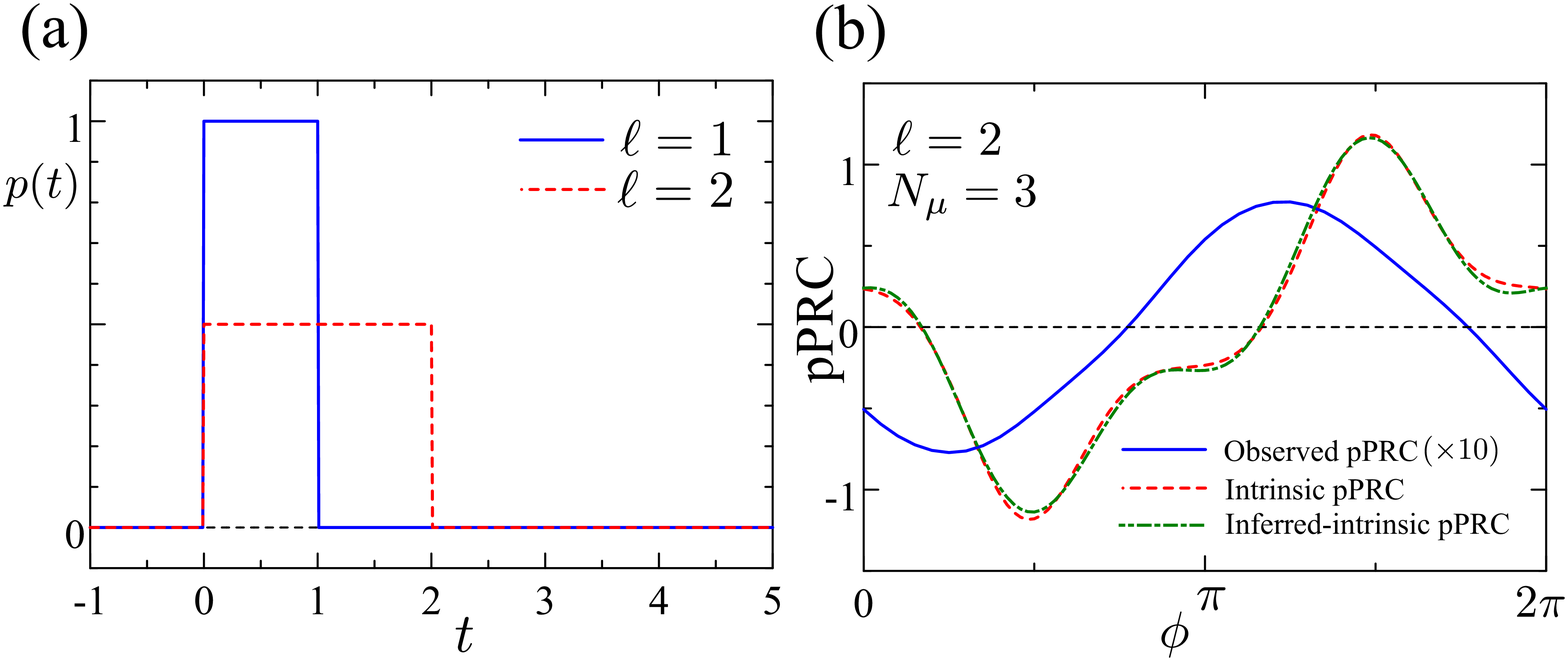}
\par\end{centering}

\caption{(a) Squared pulses $p(t;t_{s}=0)$ with the duration $\ell=1$ (solid
line) and $\ell=2$ (dashed line). (b) pPRCs of the oscillator model
of Eqs.~\eqref{eq:FHNx}--\eqref{eq:FHNy}: observed pPRC (solid
line; the magnitude is multiplied by 10) which is observed with $\ell=2$
and $\chi=0.1$, intrinsic pPRC (dashed line) and inferred intrinsic
pPRC (dot-dashed line) by Eq.~\eqref{eq:trans1}. \label{fig:PRC_inference}}
\end{figure}

pPRCs discussed in the main text are the intrinsic pPRCs, which govern
internal phase dynamics of clocks. To observe the intrinsic pPRCs
experimentally, short light pulses are applied to organisms to measure
the phase shift induced by the pulses. Here we study a relation between
intrinsic and observed pPRCs. We assume that a parameter $\rho$ in
clocks is perturbed $\rho+d\rho$ with $d\rho=\chi p(t)$ where the
phase $\phi\in[0,2\pi)$ obeys the same equation as Eq.~\eqref{eq:dphidt}.
Thus we use results of the perturbation expansion represented by Eqs.~\eqref{eq:zeroth_order}
and \eqref{eq:first_order}. For an input signal $p(t)$, we employed
the squared pulse defined by 
\begin{equation}
p(t;t_{s})=\frac{1}{\ell}H(t-t_{s})H(\ell+t_{s}-t),\label{eq:input_signal}
\end{equation}
where $H(t)$ is the Heaviside step function defined by $H(t)=1$
for $t\ge0$ and $H(t)=0$ for $t<0$, $t_{s}$ is onset time of the
squared pulse and $\ell$ is the pulse duration (we express $p(t)=p(t;t_{s})$
to represent explicit dependence on $t_{s}$). We plotted Eq.~\eqref{eq:input_signal}
for two $\ell$ cases: $\ell=1$ (solid line) and $\ell=2$ (dashed
line) in Fig.~\ref{fig:PRC_inference}(a). With Eq.~\eqref{eq:input_signal},
Eq.~\eqref{eq:phi1} is specifically represented by 
\begin{equation}
\phi_{1}(t_{e})=\int_{0}^{t_{e}}Z(\Omega t)p(t;t_{s})dt=\frac{1}{\ell}\int_{t_{s}}^{t_{s}+\ell}Z(\Omega t)dt.\label{eq:phi1_te}
\end{equation}
Here we expand $Z(\phi)$ by the Fourier series up to $N_{\mu}$th
order
\begin{equation}
Z(\phi)=\sum_{\mu=-N_{\mu}}^{N_{\mu}}c_{\mu}\exp(\mathrm{i}\mu\phi),\label{eq:Z_Fourier}
\end{equation}
where $c_{\mu}$ is an expansion coefficient. Substituting Eq.~\eqref{eq:Z_Fourier}
into Eq.~\eqref{eq:phi1_te}, we have
\[
\phi_{1}(t)=\sum_{\mu=-N_{\mu}}^{N_{\mu}}\frac{\mathrm{i}c_{\mu}}{\ell\mu\Omega}\left\{ 1-\exp\left(\mathrm{i}\mu\Omega\ell\right)\right\} \exp\left(\mathrm{i}\mu\Omega t_{s}\right).
\]
Let $\tilde{Z}(\phi;\ell)$ be an observed pPRC by the squared pulse
with the duration $\ell$. Because the observed pPRC as a function
of onset phase of the squared-pulse is quantified by the phase shift
induced by the squared-pulse, the observed pPRC $\tilde{Z}(\phi;\ell)$
can be expressed by
\begin{equation}
\tilde{Z}(\phi;\ell)=\chi\sum_{\mu=-N_{\mu}}^{N_{\mu}}\frac{\mathrm{i}c_{\mu}}{\ell\mu\Omega}\exp(\mathrm{i}\mu\phi)\left\{ 1-\exp(\mathrm{i}\mu\Omega\ell)\right\} .\label{eq:Ztilde_ck}
\end{equation}
We also expand $\tilde{Z}(\phi;\ell)$ with respect to the Fourier
series: 
\begin{equation}
\tilde{Z}(\phi;\ell)=\sum_{\mu=-N_{\mu}}^{N_{\mu}}\tilde{c}_{\mu}\exp(\mathrm{i}\mu\phi),\label{eq:Ztilde_Fourier}
\end{equation}
where $\tilde{c}_{\mu}$ is an expansion coefficient. Equating Eqs.~\eqref{eq:Ztilde_ck}
and \eqref{eq:Ztilde_Fourier}, we can represent $c_{\mu}$ by
\begin{equation}
c_{\mu}=\begin{cases}
{\displaystyle -\frac{1}{\chi}\frac{\mathrm{i}\tilde{c}_{\mu}\mu\Omega\ell}{1-\exp(\mathrm{i}\mu\Omega\ell)}} & \mu\ne0\\
{\displaystyle \frac{\tilde{c}_{\mu}}{\chi}} & \mu=0
\end{cases},\label{eq:trans1}
\end{equation}
which shows that the observed pPRC agrees with the intrinsic one (except
for scaling) only when the squared pulse is the $\delta$-peaked function
($\ell=0$). For $\ell>0$, it is impossible to completely restore
the intrinsic pPRC from the observation, because higher order harmonics
in the intrinsic pPRC are masked by the duration ($\mu$th order harmonics
where $|\mu|$ is sufficiently smaller than $T/\ell$ can be properly
restored). 

In order to check the reliability of the inference method, we applied
to a simple oscillator model 
\begin{align}
\frac{dx}{dt} & =x-x^{3}-y+\rho,\label{eq:FHNx}\\
\frac{dy}{dt} & =x,\label{eq:FHNy}
\end{align}
where $\rho$ is a light sensitive parameter (Eq.~\eqref{eq:FHNx}
corresponds to an additive case). The intrinsic pPRC of the oscillator
is described by dashed line in Fig.~\ref{fig:PRC_inference}(b).
For test data, we artificially generated an observed pPRC which is
the phase difference $\Delta\phi$ caused by the squared-pulse with
the duration $\ell$. We applied $\rho\rightarrow\rho+d\rho$ to Eq.~\eqref{eq:FHNx}
where $\rho=0$ and $d\rho=\chi p(t;t_{s})$ ($\ell=2$ and $\chi=0.1$)
and the phase difference $\Delta\phi$ as a function
of onset phase of the pulse is used as the observed pPRC which is
plotted in Fig.~\ref{fig:PRC_inference}(b) (solid line; the magnitude
is multiplied by 10). The dot-dashed line in Fig.~\ref{fig:PRC_inference}(b)
shows an inferred intrinsic pPRC through Eq.~\eqref{eq:trans1} which
is calculated with $N_{\mu}=3$ order approximation (i.e. we approximated
the observed pPRC with the third order Fourier series and applied
Eq.~\eqref{eq:trans1}). We see excellent agreement between the inferred
(dot-dashed line) and true (dashed line) intrinsic pPRCs, which verifies
the reliability of the inference method.